
\input phyzzx
\tolerance=1000
\voffset=-0.3cm
\hoffset=1cm
\sequentialequations
\def\rl{\rightline}

\def\r#1{$\bf#1$}

\def\t1{{\tilde 1}}

\def\DVN{D. V. Nanopoulos}

\def\NPB#1#2#3{Nucl. Phys. B {\bf#1} (19#2) #3}
\def\PLB#1#2#3{Phys. Lett. B {\bf#1} (19#2) #3}
\def\PRD#1#2#3{Phys. Rev. D {\bf#1} (19#2) #3}

\def\PRT#1#2#3{Phys. Rep. {\bf#1} (19#2) #3}

\def\IJMP#1#2#3{Int. J. Mod. Phys. A {\bf#1} (19#2) #3}

\def\l{\langle}
\def\r{\rangle}

\REF\WIT{E. Witten, \NPB{202}{82}{253}.}
\REF\NR{L. Girardello and M. T. Grisaru, \NPB{194}{82}{65}.}
\REF\SUGRA{H.P. Nilles, \PRT{110}{84}{1}; \DVN~ and Lahanas, \PRT{145}{87}{1}.}
\REF\GSW{M. Green, J. Schwarz and E. Witten, Superstring Theory Vols. 1,2
(Cambridge University Press, Cambridge, 1987).}
\REF\GCM{H. P. Nilles, \PLB{115}{82}{193};
J. P. Deredinger, L. E. Ibanez and H. P. Nilles, \PLB{155}{85}{65};
M. Dine, R. Rohm, N. Seiberg and E. Witten, \PLB{156}{85}{55}.}
\REF\FER{S. Ferrara, L. Girardello and H. P. Nilles, \PLB{125}{83}{457}.}
\REF\WNP{S. Ferrara, N. Magnoli, T. Taylor and G. Veneziano, \PLB{245}{90}
{409}.}
\REF\LT{D. Lust and T. Taylor, \PLB{253}{91}{335}.}
\REF\CAS{J. A. Casas, Z. Lalak, C. Munoz and G. G. Ross, \NPB{347}{90}{243}.}
\REF\IBA{A. Font, L. Ibanez, D. Lust and F. Quevedo, \PLB{245}{90}{401}.}
\REF\CVET{M. Cvetic, A. Font,L. Ibanez, D. Lust and F. Quevedo, \NPB{361}{91}
{194}.}
\REF\NIL{H. P. Nilles and M. Olechowsky, \PLB{248}{90}{268}; P. Binetruy and
M. Gaillard, \PLB{253}{91}{119}.}
\REF\TSD{A. Giveon, M. Porrati and E. Rabinovici, Phys. Rep. {\bf 244} (1994)
77 and references therein.}
\REF\DIL{E. Witten, \PLB{155}{85}{151}.}
\REF\CAR{N. V. Krasnikov, \PLB{193}{87}{37};
B. de Carlos, J. Casas and C. Munoz, \NPB{399}{93}{623}.}
\REF\FFF{I. Antoniadis, C. Bachas, and C. Kounnas, \NPB{289}{87}{87};
I. Antoniadis and C. Bachas, \NPB{298}{88}{586};
H. Kawai, D.C. Lewellen, and S.H.-H. Tye, Nucl. Phys. B {\bf 288} (1987) 1.}
\REF\EA{A. E. Faraggi and E. Halyo, preprint IASSNS-HEP-94/17, hep-ph/9405223.}
\REF\EDI{E. Halyo, \PLB{343}{95}{161}.}
\REF\CKM{A. E. Faraggi and E. Halyo, \PLB{307}{93}{305}; \NPB{416}{94}{63};
E. Halyo, \PLB{332}{94}{66}.}
\REF\SQCD{D. Amati, K. Konishi, Y. Meurice, G. C. Rossi and G. Veneziano,
\PRT{162}{88}{169}.}
\REF\MOD{E. Halyo, \NPB{438}{95}{138}.}
\REF\KLN{S. Kalara, J. L. Lopez and D. V. Nanopoulos, \PLB{245}{91}{421};
\NPB{353}{91}{650}.}
\REF\RP{E. Halyo, Mod. Phys. Lett. {\bf A9} (1994) 1415.}
\REF\MUP{E. Halyo, \NPB{424}{94}{39}.}
\REF\NANO{J. L. Lopez, D. V. Nanopoulos and K. Yuan, \PRD{50}{94}{4060}.}
\REF\FONT{L. Ibanez, W. Lerche, D. Lust and S. Theisen, \NPB{352}{91}{435}.}
\REF\DSW{M. Dine, N. Seiberg and E. Witten, \NPB{289}{87}{589};
J. J. Atick, L. J. Dixon and A. Sen, \NPB{292}{87}{109};
	 S. Cecotti, S. Ferrara and M. Villasante, \IJMP{2}{87}{1839}.}
\REF\DINE{M. Dine and A. E. Nelson, \PRD{47}{93}{1277}; M. Dine, A. E. Nelson
and Y. Shirman, \PRD{51}{95}{1362}.}
\REF\THR{V. Kaplunovsky, \NPB{307}{88}{145}; L. Dixon, V. Kaplunovsky and
J. Louis, \NPB{355}{91}{649}; I. Antoniadis, J. Ellis, R. Lacaze and D. V.
Nanopulos, \PLB{268}{91}{188}.}
\REF\SSB{V. Kaplunovsky and J. Louis, \PLB{306}{93}{269}; A. Brignole,
L. Ibanez and C. Munoz, \NPB{422}{94}{125}.}
\REF\KAH{J. Lopez and D. Nanopoulos, preprint CTP-TAMU-60/94, hep-ph 9412332.}

\singlespace
\rl{WIS--95/17/MAR--PH}
\rl{\today}
\rl{T}
\pagenumber=0
\normalspace
\smallskip
\titlestyle{\bf{Hidden Matter Condensation Effects on
Supersymmetry Breaking }}
\smallskip
\author{Edi Halyo{\footnote*{e--mail address: jphalyo@weizmann.bitnet}}}
\smallskip
\centerline {Department of Particle Physics}
\centerline {Weizmann Institute of Science}
\centerline {Rehovot 76100, Israel}
\vskip 2 cm
\titlestyle{\bf ABSTRACT}

We investigate the effects of hidden matter condensation on supersymmetry
breaking in supergravity models derived from free fermionic strings.
We find that the minimum of the effective potential in the modulus direction
depends strongly on only one parameter which is fixed by the hidden sector.
For nonpositive values of the parameter the potential is
unstable which constrains realistic models severely.
For positive and decreasing values which correspond to more and/or lighter
hidden matter, $T_R$ increases whereas $T_I$ is periodic and depends on the
parameter very weakly. Supersymmetry can be broken in the matter direction
with a stable vacuum only if the fields which give mass to the hidden matter
are light and have modulus independent Kahler potentials. Then, for a wide
range of
model parameters, supersymmetry is mainly broken by hidden matter condensation
in the matter direction rather than by hidden gaugino condensation in the
modulus direction.

\singlespace
\vskip 0.5cm
\endpage
\normalspace

\centerline{\bf 1. Introduction}

There is growing though circumstantial experimental evidence for believing
that supersymmetry (SUSY) is a true symmetry of Nature. If SUSY
exisits, it must be broken which can only happen nonperturbatively[\WIT]
due to the well-known nonrenormalization theorems[\NR]. The best candidate for
dynamical SUSY breaking seems to be condensation effects in the hidden sectors
of supergravity (SUGRA)[\SUGRA] or superstring[\GSW] models.

SUSY breaking by gaugino condensation[\GCM] in the hidden sectors of SUGRA or
string derived SUGRA models have been extensively examined in the literature
[\FER-\NIL]. In this scenario, when the
hidden gauge group becomes strong at a hierarchically small scale $\Lambda_H$
compared with the Planck scale $M_{P}$, gaugino condensates which break
SUSY form. The effects of SUSY
breaking in the hidden sector are communicated to the observable sector by
gravity and possibly by nonrenormalizable terms in the superpotential which are
proportional to inverse powers of $M_P$. The effective
nonperturbative superpotential can be obtained from the symmetries
of the underlying gauge theory, i.e. by satisfying the anomalous and
nonanomalous Ward identities[\WNP,\LT]. On the other hand, one can find the
effective scalar potential by simply substituting the VEV of the
gaugino condensate into the SUGRA Lagrangian[\FER].
A very important ingredient in this scenario is target space duality which
is a symmetry of the string to all orders in perturbation theory and also
assumed to hold nonperturbatively[\TSD]. As a result, the nonperturbative
effective superpotential has to be invariant under target space duality,
a property which restricts its possible moduli dependence severely.
The dilaton dependence of the nonperturbative superpotential is determined
by the running coupling constant which is a function of the dilaton[\DIL].

For a hidden sector without matter, hidden sector gaugino condensation
gives an effective scalar potential which has minima or vacua at
$ T_R  \sim 1.23$ and its dual value $T_R \sim 0.81$. On the other hand,
the scalar
potential is not stable in the dilaton direction resulting in $S_R \to \infty$
[\IBA]. The dilaton potential can be
stabilized either by adding a dilaton independent term to the superpotential
[\IBA] or by having more than one hidden gauge group[\CAS]. These vacua
break SUSY in the modulus but not the dilaton direction, i.e. $\l F_T \r
\not=0$
but generically $\l F_S \r =0$.  In addition, in all
versions of this scenario, the cosmological constant is nonvanishing (and
negative).

In string or string derived SUGRA models the generic situation is a
hidden sector with matter and not the pure gauge case. The effect of hidden
matter on SUSY breaking must be taken into account
unless all hidden matter multiplets are heavy and decouple at the
condensation scale. This case has been considered
assuming that SUSY is not broken in the matter direction in Refs. [\LT,\CAR].
There, it was argued that the presence of hidden matter does not change the
results of the pure gauge case significantly. In this paper, we will show that
this is not so at least in SUGRA derived from free fermionic
superstrings[\FFF] .
The presence of hidden matter with nonzero mass (as required to have
a stable vacuum)
has two effects. First, it modifies the nonperturbative superpotential
of the pure gauge case in a well--known way[\LT]. Second, as was shown in Refs.
[\EA,\EDI] it may result in SUSY breaking in the matter direction if the
fields which give mass to hidden matter do not decouple at the condensation
scale $\Lambda_H$.

In Ref. [\EDI] the effects of hidden matter condensation on SUSY breaking
were investigated in a generic SUGRA model derived from free
fermionic superstrings.
The F--terms in the overall modulus and matter directions were obtained from
the effective superpotential and compared with each other without
finding the vacuum (or the scalar, dilaton and moduli VEVs) by minimizing
the effective scalar potential.
It was shown that SUSY can mainly be broken by hidden matter
condensation in the matter direction rather than by hidden gaugino condensation
in the modulus direction. Whether the former or the latter is dominant depends
on the parameters of the string or SUGRA model such as the hidden gauge group,
hidden matter content and their masses and the vacuum, i.e. the VEVs for the
dilaton, moduli and scalar fields which are fixed dynamically.
In this paper, we extend our previous work by examining the effect of
hidden matter condensation on SUSY breaking with or without matter F--terms
again in a generic SUGRA model derived from free fermionic strings.
We make no attempt to solve the dilaton stability or the cosmological
constant problems. We assume that the dilaton VEV is stabilized at $S \sim 1/2$
by some mechanism which may be either of the two mentioned above.
We perform our analysis by obtaining the effective scalar potential for the
different cases we examine and minimizing it numerically.

We find that when SUSY is not broken in the matter direction,
the location of the minima, maxima and saddle points in the modulus direction
depend mainly on the parameter $d^{\prime}=
(6N-2M-t)/4 \pi N$. Here $N,M$ and $t$ are the hidden gauge group (assumed
to be $SU(N)$), the number of hidden matter multiplets in the fundamental
representation and the power of $\eta(T)$ in the determinant of the hidden
matter mass matrix respectively. In this case,
$T_R$ at the maxima and saddle points increase
with decreasing $d^{\prime}$ which corresponds to more and /or lighter
hidden matter. $T_R$ at the minima behave the same way except that in addition
there appear new
solutions with small (i.e. $<0.2$) $T_R$ for small values (i.e. $<1/7$)
of $d^{\prime}$. For all critical
points $T_I$ is periodic and (almost) independent of $d^{\prime}$.

When SUSY is broken in the matter and modulus directions, i.e.
there is a nonvanishing matter F--term $F_{\phi_i}$, the results depend
on the Kahler potential of the matter field $\phi_i$. If
$K(\phi_i,\phi_i^{\dagger})$ depends on the modulus, there is no stable
minimum in the $T_R$ direction. On the other hand, if the Kahler potential
of $\phi_i$ is canonical there are stable vacua for all positive values of
$d^{\prime}$. The stability of the vacuum requires that if SUSY is broken
in the matter direction, the same matter fields must have canonical Kahler
potentials. In this case,
$T_R,T_I$ at the minima behave similarly to the $F_{\phi_i}=0$
case but now there are no new minima
at small $T_R$ for small $d^{\prime}$. This is a crucial difference because
it is exactly at these points that $F_T>F_{\phi_i}$ whereas for all minima
with large $T_R$, $F_{\phi_i}>F_T$ for most of the model parameter space.
Thus, we conclude that when $F_{\phi_i} \not =0$,
the dominant SUSY breaking mechanism is hidden matter
condensation in the matter direction rather than hidden gaugino condensation
in the modulus direction.
We also find that, whether matter F--terms vanish or not, for
$d^{\prime} \leq 0$, there is no minimum in the $T_R$ direction , i.e.
$T_R \to \infty$ which gives $F_T=0$ (and $F_{\phi_i}=0$). Requiring a stable
vacuum in the modulus direction constrains the hidden sectors of
possible realistic models severely.

The paper is organized as follows. In Section 2, we give the features which are
common to all realistic SUGRA models derived from free fermionic strings.
These include the matter and moduli content, the superpotential,
the Kahler potential
and the supersymmetric vacuum around the Planck scale. In Section 3, we review
SUSY breaking by hidden gaugino condensation in the pure gauge case.
The effects of hidden matter condensation without a matter F--term
are considered in Section 4. We minimize
the effective scalar potential numerically and find that the presence of hidden
matter modifies the pure gauge case results significantly.
In Section 5, we examine hidden matter condensation effects in the
presence of a nonvanishing F--term
in the matter direction. We consider two cases: matter with
a modulus dependent Kahler potential and with a canonical one.
We find the conditions under which the matter or modulus
F--term dominates SUSY breaking. Section 6 contains a discussion of our
results and our conclusions.

\bigskip
\centerline{\bf 2. Supergravity models derived from free fermionic strings}

The low--energy effective field theory limit of superstrings is given by
$N=1$ SUGRA models with a gauge group and field content fixed by the
underlying string. The string derived SUGRA model is defined in addition by
three functions: the Kahler potential $K$, the superpotential $W$ and the gauge
function $f$[\SUGRA]. SUGRA derived from free fermionic strings have some
generic features which we outline below. These can also be considered as
assumptions about the string models which we examine in this paper.
We consider a SUGRA model derived from a free fermionic superstring
[\FFF] with the following properties:

a) The spectrum of the SUGRA model which is given by the massless spectrum
of the superstring is divided into three sectors. The first one is the
observable sector which contains states with charges under the Standard Model
gauge group. The second one, the hidden sector, contains singlets of
the Standard Model group which are multiplets of the hidden gauge group.
These two sectors are connected only by nonrenormalizable
terms in the superpotential, gravity and gauged $U(1)$s which are broken
around the the Planck scale. Therefore, once the gauged $U(1)$s are
broken the two sectors are
connected only by interactions proportional to inverse powers of $M_P$.
The third sector generically contains a large number of matter fields
($\phi_i$)
which are Standard Model and hidden gauge group singlets. $\phi_i$ are
connected to the observable and hidden states only through gauge $U(1)$s
(in addition
to gravity and nonrenormalizable interactions which are proportional to
inverse powers of $M_P$). Therefore, $\phi_i$ behave effectively as hidden
matter once the $U(1)$s are broken. Throughout the paper we call the fields
$\phi_i$ matter (not observable or hidden). It is the F--terms of these
$\phi_i$ that we are interested in when we examine SUSY breaking in the matter
direction.

b) The hidden sector contains one (or more) $SU(N)$ (or other nonabelian gauge)
group(s) with $M$ copies of matter ($h_i, \bar h_i$) in the vector
representations $N+\bar N$. In the following, we consider only the one gauge
group case since in realistic models part of the hidden gauge group must
be broken by VEVs which are essential for obtaining CKM mixing[\CKM]. In
any case our results are not changed
by the introduction of additional hidden gauge groups with matter. The case of
multiple hidden gauge groups has been extensively examined in Ref. (\CAR).
The net
effect of additional hidden gauge groups is to stabilize the dilaton potential
which we assume in the following. The hidden matter states obtain masses
from nonrenormalizable terms, $W_n$, of the type given in Eq. (2) below.
This is essential since a supersymmetric gauge theory with massless matter
does not have a well--defined vacuum[\SQCD]. As a result of the
nonrenormalizable terms, the hidden matter mass
matrix is nonsingular and the model has a stable vacuum. In addition,
$M<3N$ so that the hidden gauge group
is asymptotically free and condenses at the scale $\Lambda_H \sim M_P
exp(8 \pi^2/bg^2)$ where $b=M-3N$.

c) Realistic free fermionic strings generically have a number of untwisted
moduli in their massless spectrum. These show up in the low--energy SUGRA
model as fields which do not appear in the superpotential to any order in
perturbation theory. One moduli always present in all string models is the
dilaton. The exact type and number of the other untwisted moduli depend
on the boundary conditions for the internal fermions and are model dependent
[\MOD]. In realistic free fermionic string models one can have up to three $T$
type and three $U$ type moduli, one pair for each compactified torus (sector).
In the following we will assume to have only one untwisted modulus which
is the overall modulus $T$ for simplicity. Thus, we will deal with
target space duality under only the overall modulus $T$.
It is straightforward to generalize
this case to the one with any number of untwisted moduli of either $T$ or $U$
type. On the other hand, there are free fermionic strings for which some or
all tori (sectors) do not have any moduli.
Matter fields arising from these sectors have
canonical Kahler potentials which do not depend on moduli.

d) The superpotential is given by
$$W=W_{3,obs}+W_{3,hid}+W_n+W_{np}, \eqno(1)$$
where the cubic superpotential $W_3$ is divided into two parts: one which
contains only the observable states and the other only the hidden states.
$W_n$ gives
the nonrenormalizable terms ($n>3$) in the superpotential and $W_{np}$ gives
the nonperturbative contributions due to gaugino and matter condensation in
the hidden sector. Due to the supersymmetric nonrenormalization theorems[\NR],
the only correction to the string tree level superpotential is $W_{np}$.
Note that there are no renormalizable interactions between observable and
hidden
matter arising from the superpotential. We assume that the same is true also
for $\phi_i$ and the Standard Model states.
We also take the gauge function $f_{\alpha \beta}=S \delta_{\alpha \beta}$
at the string tree level[\DIL]. Neglecting the string one loop corrections to
$f_{\alpha \beta}$ do not change our results qualitatively.

e) The nonrenormalizable (order $n>3$) terms in the superpotential
are generically of the form
$$W_n=c_n g h_i \bar h_j \phi_{j_1} \phi_{j_2} \ldots \phi_{j_{n-2}}
\eta(T)^{2n-6} M_v^{3-n}, \eqno(2)$$
and are obtained from the world--sheet correlators
$$A_n \sim \l V_1^f V_2^f V_3^b \ldots V_n^b \r, \eqno(3)$$
which satisfy all the selection rules due to the local and global charges and
Ising model operators as given by Ref. (\KLN). In Eq. (2), $c_n$
are calculable numerical coefficients of $O(1)$ and
$\eta(T)=e^{-\pi T/12} \prod_n (1-e^{-2\pi n T})$ is the Dedekind eta function.
In free fermionic strings, modular weights of matter fields under the
overall modulus $T$, are given by the sum $\sum_{i=1}^3 Q_{\ell_i}$ where
$Q_{\ell_i}$ give their R charges[\RP]. A generic feature of free
fermionic strings is that all matter fields $\phi_i$ and $h_i$ have modular
weights $-1$. Thus, the cubic superpotential $W_3$ is automatically target
space modular
invariant. Nonrenormalizable terms $W_n$ are rendered modular invariant by
multiplying them by the required powers of $\eta(T)$ which has a modular
weight of $1/2$. In Eq. (2) the powers of $\eta(T)$ and $M_v$ ($\sim M_P$
to be defined later) are
such that the term $W_n$ has modular
weight $-3$ and dimension $3$ as dictated by dimensional analysis and
target space modular invariance.
Note that these terms contain both observable and at least a pair of
hidden sector states. Once
the fields $\phi_i$ get VEVs (in order to have a supersymmetric vacuum at $M_P$
as a result of the anomalous D--term as we will see below), they give
masses to the hidden states $h_i,
\bar h_i$. Consequently, all the $n>3$ terms of the type given by Eq. (2)
can be seen as hidden matter mass terms.
(In general, there can also be terms of the form $c_n \phi_{i_1} \phi_{i_2}
\ldots \phi_{i_n}$, i.e. nonrenormalizable terms with only observable fields.
These vanish in standard--like models [\EA] and we assume that they are not
present in the following. Elimination of these terms is closely related to
discrete symmetries which protect light quark masses[\MUP].
If they do exist, they may destabilize the SUSY
vacuum and break SUSY at very large scales which is phenomenologically a
disaster.)

f) The Kahler potential (at tree level) is given by (for $\phi_i <<T$)
$$K(S, S^\dagger, T, T^\dagger, \phi_i,  \phi_i^\dagger)=-log(S+S^\dagger)-
3log(T+ T^\dagger)-\sum_i (T+ T^\dagger)^{n_i}\phi_i \phi_i^\dagger, \eqno(4)$$
where $S,T$ and $\phi_i$ are the dilaton, (overall) modulus and matter fields
respectively and $n_i$ is the modular weight (under $T$) of the matter
field $\phi_i$. There are also models in which some sectors do not have any
moduli. Matter fields coming from these sectors have canonical Kahler
potentials which do not depend on the modulus. The presence of such matter
fields will be crucial for stablizing the scalar potential with matter F--terms
in Section 5.
The modulus and matter fields in Eq. (4) are in the ``supergravity basis"
and are related to the massless string states by well--known transformations
[\NANO,\FONT].

g) The string vacuum is supersymmetric at the Planck scale, $M_P$ and at the
level of the cubic superpotential. This is
guaranteed by satisfying the F and D constraints obtained from the cubic
superpotential $W_3$ (which is trilinear in $\phi_i$ and $h_i$) and the local
charges of the states.
As we saw above,
all nonrenormalizable terms in the superpotential, $W_n$, contain
hidden matter bilinears. As a result, $W_3$ does not get any higher
order corrections as long as the hidden gauge group does not condense at
$\Lambda_H<<M_P$ and $W_3$ is the exact superpotential until hidden
sector condensation. The set of F and D
constraints is given by the following set of equations [\EA]:
$$\eqalignno{&D_A=\sum_i Q^A_i \vert \l \phi_i \r\vert^2={-g^2
\over 192\pi^2}Tr(Q_A) {1\over {2\alpha^{\prime}}}, &(5a) \cr
&D^j=\sum_i Q^j_i \vert \l \phi_i \r\vert^2=0 ,&(5b) \cr
&\l W_3 \r=\l{\partial W_3\over \partial \phi_i} \r=0 , &(5c) \cr}$$
where $\phi_i$ are the matter fields and $Q^j_i$ are their local charges.
$\alpha^{\prime}$ is the string tension
given by $(2\alpha^{\prime})^{-1}=g^2M_P^2/32\pi=g^2M_v^2$
and $Tr(Q_A)\sim 100$ generically in realistic string models. Eq. (5a) is the
D constraint for the anomalous $U(1)_A$ which is another generic
feature of free fermionic string models [\DSW]. Note that the anomalous D--term
arises at the string one loop level and therefore contains a factor of
$g^2=1/4(S+S^{\dagger})$.
We see that some Standard Model singlet scalars must get Planck scale VEVs
of $O(M_v/10)$ in order to satisfy Eq. (5a)
and preserve SUSY around the Planck scale. Then, due to the other F and D
constraints most of the other SM singlet scalars also obtain VEVs
of $O(M_v/10)$. In this manner, all gauge $U(1)$s which connect
$\phi_i$ and $h_i,\bar h_i$ to the Standard Model states
are broken spontaneously at the high
scale $O(M_v/10)$. In addition, the scalar VEVs
break target space duality spontaneously since they carry modular weights.
These corrections to $W_3$ when they become nonzero, (i.e.
when hidden matter condensates $\Pi_{ij}=h_i \bar h_j$ form) modify
the cubic level F constraints in Eq. (5a) and may destabilize the original
SUSY vacuum in the matter direction as was shown in Ref. (\EA).

\bigskip
\centerline{\bf 3. Hidden sector gaugino condensation}

The leading candidate for SUSY breaking in string derived SUGRA is hidden
sector gaugino condensation[\GCM]. In this section, we review the simplest
possibility which is gaugino condensation in a hidden sector with a pure gauge
group i.e. no hidden matter. As mentioned previously,
realistic string models generically
contain hidden matter in vectorlike representations. This more complicated
case will be discussed in the following sections. Our
purpose in reviewing the pure gauge case is to introduce the basic concepts
and our notation.

In this scenario, due to the running of the coupling constant, the hidden
gauge group condenses around the scale $\Lambda_H \sim M_P Exp(8 \pi^2/bg^2)$,
resulting in a gaugino
condensate. The nonperturbative effective superpotential for the gaugino
condensate $Y^3$
can be obtained from the symmetries (Ward identities) of the underlying gauge
theory to be[\WNP,\LT]
$$W_{np}={1 \over {32\pi^2}}Y^3 log\{exp(32\pi^2S)[c \eta(T)]^{6N} Y^{3N} \}
, \eqno (6)$$
where $c$ is a constant. $W_{np}$ has modular weight $-3$ as required since
$Y^3$ and $S$ have modular weights $-3$ and $0$ respectively. All the fields
which appear in $W_{np}$ are scaled by $M_v$.
The composite gaugino condensate superfield $Y^3$
can be integrated out by taking the flat limit $M_P \to \infty$
at which gravity decouples. In this limit, SUGRA reduces to
global SUSY whose vacuum is given by
$${\partial W_{tot} \over {\partial Y}}=0. \eqno(7)$$
The solution to the above equation gives the gaugino condensate in terms of
$S$ and $T$
$${1 \over {32\pi^2}}Y^3=(32\pi^2e)^{-1}[c \eta(T)]^{-6}
exp(-32\pi^2S/N), \eqno (8)$$
resulting in the nonperturbative superpotential
$$W_{np}(S,T)=\Omega(S) h(T) , \eqno(9)$$
with
$$\eqalignno{&\Omega(S)=-N exp(-32\pi^2S/N), &(10a) \cr
             &h(T)=(32\pi^2e)^{-1}[c \eta(T)]^{-6}. &(10b)}$$
The effective scalar potential due to $W_{np}$ is given by
$$V=|F_S|^2 G_{SS^\dagger}^{-1}+|F_T|^2 G_{TT^\dagger}^{-1}-3e^K|W|^2,
\eqno(11)$$
where $G=K+log|W|^2$, $W=W_{np}$ ($W_3=0$ in vacuum from Eq. (5)) and the
F--terms are
$$F_k=e^{K/2}(W_k+K_kW), \eqno(12)$$
for $k=S,T$. Using the above formula we find (from now on we
use the notation $S=S_R+iS_I$ and $T=T_R+iT_I$)
$$F_S={1 \over {(2S_R)^{1/2}(2T_R)^{3/2}}} h(T)
\left(\Omega_S-{\Omega \over {2S_R}} \right), \eqno (13)$$
and
$$F_T={1 \over {(2S_R)^{1/2} (2T_R)^{3/2}}} \Omega(S)
\left(h_T-{3h \over {2T_R}} \right). \eqno(14)$$
Substituting the above F--terms into Eq. (11) gives the scalar potential
$$V={1\over {16S_R T_R^3 |\eta(T)|^{12}}} \left\{|2S_R \Omega_S-\Omega|^2+
3|\Omega|^2 \left({T_R^2 \over \pi^2}|\hat G_2|^2-1 \right)\right\},
\eqno(15)$$
where $\hat G_2=G_2-\pi/T_R$ and
$G_2$ is the second Eisenstein function given by
$$G_2(T)={\pi^2 \over 3}-8 \pi^2\sum_n \sigma_1(n)e^{-2\pi n T}. \eqno(16)$$
$\sigma_1(n)$ is the sum of the divisors of $n$ and $G_2(T)$ arises due to
$${\partial \eta(T) \over \partial T}=-{\eta(T) \over {4\pi}}G_2(T).
\eqno(17)$$
$\hat G_2(T)$ is a regularized version of $G_2(T)$
which has modular weight $-2$ (in contrast $G_2(T)$ does not have a
well--defined modular weight)[\CVET]. The vacuum is obtained by minimizing
the scalar potential with repect to $S$ and $T$.
If any of the F--terms given by Eqs. (13) and (14) are nonzero in the vacuum,
SUSY is broken spontaneously.
It is well--known that the condition for a minimum in the $S$ direction
is given by[\IBA]
$$S_R \Omega_S-\Omega=0. \eqno(18)$$
Note that the minimum in the $S$ direction does not depend on the modulus $T$.
With $\Omega(S)$ given by Eq. (10a) one finds that there is no (finite)
minimum or (stable) vacuum since the solution to Eq. (18) requires
$S_R \to \infty$. This is the dilaton
stability problem and we will not try to solve it in this paper. It has
been noted that
the dilaton potential can be stabilized with a realistic dilaton VEV, i.e.
$S_R \sim 1/2$ either by adding a constant term to $\Omega(S)$[\IBA] or by
having more than one hidden gauge group[\CAR]. We stress that the condition for
the minimum in the $S$ direction automatically insures $F_S=0$ which we
will assume to hold in the following.

The minimization in the modulus direction gives
$$\eqalignno{{\partial V \over \partial T}&={3 \over {32 \pi S_R T_R^3}}
{1 \over {|\eta(T)|^{12}}}
\{\hat G_2[|2S_R \Omega_S-\Omega|^2+3|\Omega|^2 \left({T_R^2 \over \pi^2}
|\hat G_2|^2-1\right)] \cr
&+{T_R \over \pi}|\Omega|^2[2|\hat G_2|^2+T_R \hat G_2^*
\hat G_{2T}+T_R \hat G^*_{2T} \hat G_2] \}=0. &(19)}$$
The first term in the curly brackets vanishes due to $F_S=0$.
{}From Eq. (19) we see that the minimum in the modulus direction is
independent of the dilaton $S$. In addition, it is also independent of the
hidden gauge group or $N$.
The critical points of the potential $V$ in Eq. (15) have been investigated
[\IBA].
There are maxima at $(T_R,T_I)=({\sqrt 3}/2,1/2+n)$
and saddle points at $(T_R,T_I)=(1,n)$ which are given by the solutions to
$\hat G_2(T)=0$. (Here $n$ is an integer.)
Both at the maxima and saddle points $F_T=0$ since $F_T \prop
\hat G_2(T)$ as it is seen from Eq. (14). The minima are given by solutions
to Eq. (19) which are not solutions of $\hat G_2(T)=0$. They are at
$(T_R,T_I)=(\sim 1.23,n)$ and its dual $(T_R,T_I)=(\sim 0.81,n)$
which give $F_T \not =0$. The maxima and saddle points
appear at the self--dual (or fixed) points of target space duality due to the
fact that $\hat G_2$ transforms covariantly
under target space duality (or has a well-defined modular weight) and
modular functions always have zeros at these fixed points. The minima, on the
other hand, are not at the fixed points of target space duality. Therefore,
target space duality which is spontaneously broken by the vacuum manifests
itself by the presence of two minima which are connected to each other by
target space duality.

\bigskip
\centerline{\bf 4. Hidden sector gaugino and matter condensation}

As mentioned in Section 2, free fermionic strings generically have hidden
sectors which contain matter $h_i, \bar h_i$ in the vectorlike representations
of the hidden
gauge group. In addition, there are generic observable matter fields $\phi_i$
which give masses to hidden matter.
In this section, we repeat the steps of the previous one taking into account
the effects of hidden matter condensation. We assume that SUSY is not broken
in the matter direction, i.e. $F_{\phi_i}=0$. The case where
$F_{\phi_i} \not=0$ will be examined in the next section.

In the presence of hidden matter, when the hidden gauge group condenses
at $\Lambda_H$, matter condensates $\Pi_{ij}=h_i \bar h_j$ form in addition to
gaugino condensates $Y^3$. The
nonperturbative effective superpotential
obtained from the Ward identities and modular invariance becomes[\WNP,\LT]
$$W_{np}={1 \over {32\pi^2}}Y^3 log\{exp(32\pi^2S)[c^{\prime} \eta(T)]^{6N-2M}
Y^{3N-3M} det \Pi \}-tr A \Pi, \eqno (20)$$
where $c^{\prime}$ is a (new) constant and $A$ is the hidden matter mass
matrix given by the $n>3$ terms in Eq. (2). $W_{np}$ has modular
weight $-3$ as required since
$A$ and $\Pi$ have modular weights $-1$ and $-2$ respectively.
The last term corresponds
to the sum of all the $n>3$ terms in Eq. (2). The observable matter fields
$\phi_i$ which give masses to hidden matter
appear only in the mass matrix $A$. In the flat limit $M_P \to \infty$,
gravity decouples and one gets a globally SUSY vacuum at which (in addition to
Eqs. (5a-c))
$${\partial W_{tot} \over {\partial Y}}={\partial W_{tot} \over
{\partial \Pi}}=0, \eqno(21)$$
where $W_{tot}=W_3+W_{np}$. We can replace $W_{tot}$ in Eq. (21) by $W_{np}$
since $W_3$ does not contain $Y^3$ or $\Pi$. The $n>3$ terms, $W_n$ which are
the hidden matter mass terms, are already included in $W_{np}$ through
$tr A \Pi$. The solutions to Eq. (21)
are used to obtain the composite fields $Y^3$ and $\Pi$ in terms of
$S,T$ and $A$
$${1 \over {32\pi^2}}Y^3=(32\pi^2e)^{M/N-1}[c \eta(T)]^{2M/N-6}[det A]^{1/N}
exp(-32\pi^2S/N), \eqno (22)$$
and
$$\Pi_{ij}={1 \over {32\pi^2}}Y^3 A^{-1}_{ij}. \eqno(23) $$
Eqs. (22) and (23) are used to eliminate the composite fields in $W_{np}$
$$W_{np}(S,T)=\Omega(S) h(T) [det A]^{1/N}, \eqno(24)$$
where
$$\eqalignno{&\Omega(S)=-N exp(-32\pi^2S/N), &(25a) \cr
             &h(T)=(32\pi^2e)^{M/N-1}[c \eta(T)]^{2M/N-6}. &(25b)}$$
$det A$ is a product of mass terms given generically by Eq. (2). Thus, without
any loss of generality, we can
assume that it has the form
$$det A=k S_R^{-r} \phi_i^{s_i} \eta(T)^t \qquad r,s,t>0, \eqno(26)$$
where the $S$ dependence is obtained from the relation $g^2=1/4S_R$
(at the string tree level and for level one Kac--Moody algebras).
The parameters $r$ and $t$ can be expressed in terms of the more
fundamental ones, hidden sector parameters $N,M$ and the order of
nonrenormalizable mass terms $n$ for a given model.
$\phi_i$ denotes any matter field which appears in $det A$ and $s_i$ is its
power. k is a constant of $O(1)$ which is given by the product of the relevant
$c_n$ in Eq. (2). In fact, this is the form of $det A$ which was obtained
from the explicit model of Ref. (\EA)
with $r=7$, $t=22$ and $s_i=1,5$ depending on the field $\phi_i$.
(In general, $det A$ is a sum of terms like that in Eq. (26).)
We see that there is a new $S$ and $T$ dependence in $W_{np}$ due to $det A$.
The new $S$ dependence does not change the results of the
previous section qualitatively. The scalar potential still has a minimum only
at $S \to \infty$ which needs to be stabilized and $F_S=0$ due to the
minimization condition. On the other hand, the new $T$ dependence leads to
qualitative and quantitative changes as we will
see below.

Now, there are two possibilities: either $\phi_i$ are heavier than
$\Lambda_H$, i.e. $m_{\phi_i}>>\Lambda_H$ and they decouple at $\Lambda_H$
or they are lighter than $\Lambda_H$, i.e. $m_{\phi_i}<\Lambda_H$ and they
remain in the spectrum. In this section, we assume the former which has two
consequences. First, since $\phi_i$ decouple at the condensation scale one can
substitute their VEVs everywhere and forget about them. Second, there is no
SUSY
breaking in the $\phi_i$ direction, i.e. $F_{\phi_i}=0$. The second case in
which $\phi_i$ are light will be examined in the next section. Then,
$\phi_i$ do not decouple and become dynamical fields like $S$ and $T$.
In both cases we assume that the hidden matter states $h_i, \bar h_i$
do not decouple from the
spectrum at $\Lambda_H$ since otherwise obviously there can only be gaugino
condensation.

In $W_{np}$ all the information about the matter condensates, $\Pi_{ij}$, and
the observable fields $\phi_i$ is contained in the term $det A$.
When $m_{\phi_i}>> \Lambda_H$ and $\phi_i$ decouple, one simply substitutes the
VEVs $\l \phi_i \r$ obtained from the solution to the F and D constraints in
$det A$. $\phi_i$ are longer dynamical fields since at the scale $\Lambda_H$
these heavy fields cannot be excited but simply sit at their VEVs. In this
sense, $\phi_i$ are similar to
the composite fields $Y^3$ and $\Pi$ which are also eliminated from $W_{np}$.
All $\phi_i$ do is to give masses to the hidden matter states $h_i, \bar h_i$
through their VEVs. As a result, in this case the only effect of matter
condensates $\Pi_{ij}$ is to change the scale of the gaugino condensate $Y^3$
through $det A$.

Using Eq. (12) we obtain for the dilaton F--term
$$F_S={e^{-{\phi_i \phi_i^\dagger}/{4 T_R}} \over {(2S_R)^{1/2}
(2T_R)^{3/2}}} h(T) [det A]^{1/N}
 \{\Omega_S-{\Omega \over {2S_R}}+\Omega (log [det A]
^{1/N})_S \}. \eqno(27)$$
The first two terms in the curly brackets are the usual ones coming from
gaugino condensation.
The last term gives the contribution of the matter condensates (through $det
A$)
to $F_S$. Assuming the above form for $det A$ we get
$${\partial (log [det A]^{1/N}) \over \partial S}=-{r \over {NS_R}}.
\eqno(28)$$
It is easy to see that this additional term can be absorbed into a redefinition
of $b$ and does not change the $F_S$ or the dilaton potential qualitatively.
Once again the dilaton has a runaway potential with $S_R \to \infty$ which
should be stabilized by some mechanism.

For the F--term in the modulus direction we find
$$\eqalignno{F_T={e^{-{\phi_i \phi_i^\dagger}/{4 T_R}} \over {(2S_R)^{1/2}
(2T_R)^{3/2}}} & \Omega(S) [det A]^{1/N}
 \{h_T-{3h \over {2T_R}} \cr
&+{\phi_i \phi_i^{\dagger} \over 4 T_R^2}h +h (log[det A]^{1/N})_T \}.&(29)}$$
As for $F_S$, the first two terms in the curly brackets  arise from gaugino
condensation whereas the last two come from matter Kahler potential $K(\phi_i,
\phi_i^{\dagger})$ and hidden matter condensation respectively.
{}From Eq. (26) for $det A$ we obtain for the last term
$${\partial (log[det A]^{1/N}) \over \partial T}= -{t \over {4 \pi N}} G_2(T).
\eqno(30)$$

Combined with the power of $\eta(T)$ in $h(T)$ in Eq. (25b), $2M/N-6$,
these two
terms modify the behavior in the modulus direction. The F--term in the modulus
direction is now given by
$$F_T={e^{-{\phi_i \phi_i^\dagger}/{4 T_R}} \over {(2S_R)^{1/2}
(2T_R)^{3/2}}} \Omega(S) [det A]^{1/N} h(T) d^{\prime}
 \left(G_2(T)-{3 \over {2T_R d^{\prime}}}+{\phi_i \phi_i^{\dagger} \over 4
T_R^2
d^{\prime}} \right), \eqno(31)$$
where $d^{\prime}=(6N-2M-t)/4 \pi N$ which gives the scalar potential
$$\eqalignno{V&={e^{-\phi_i \phi_i^{\dagger}/2T_R} \over {16S_R T_R^3
|\eta(T)|^{8 \pi d^{\prime}}}}
|[det A]^{1/N}|^2 \{|2S_R \Omega_S-\Omega-{2 \Omega r \over N}|^2 \cr
&+|\Omega|^2 \left({4 d^{\prime2}T_R^3 \over (3 T_R- \phi_i \phi_i^{\dagger})}
|G_2(T)-{3 \over {2T_R d^{\prime}}}+{\phi_i \phi_i^{\dagger} \over {4 T_R^2
d^{\prime}}}|^2-3\right)\}. &(32)} $$
We see that the effect of hidden matter condensates and their mass terms is
simply to change the function $\hat G_2(T)$ to $G_2(T)-{3/ 2T_R d^{\prime}}
+{\phi_i \phi_i^{\dagger}/ 4 T_R^2 d^{\prime}}$
where $d^{\prime}$ is fixed by the hidden gauge group ($N$),
the matter content of the hidden sector ($M$) and the hidden
mass terms ($t$) in Eq. (26). The matter VEVs $\l \phi_i \r$ are fixed by the
F and D--terms of the of the superpotential to be $\sim M/10$ at the
perturbative level. The additional nonperturbative scalar potential is much
smaller than the perturbative one and therefore cannot change the matter VEVs
by much. Any supersymmetric string vacuum contains a large number (of
$O(10)$) of VEVs and
therefore for our calculations we take $\l \phi_i \phi_i^{\dagger} \r \sim
0.2$.
We have numerically checked that our results are not sensitive to the exact
value and number of the VEVs as long as they are nonzero and in a realistic
range.

Note that, as expected, for $M=t=0$ and $\l \phi_i \r=0$, $G_2^{\prime}(T) \to
\hat G_2(T)$ and the potential in Eq. (32) reduces to the pure gauge
result given by Eq. (15).
In complete analogy with the pure gauge case, now the
behavior in the modulus direction is determined by the function
$G_2(T)-{3/ 2T_R d^{\prime}}+{\phi_i \phi_i^{\dagger}/ 4 T_R^2
d^{\prime}}$. Minimizing in the dilaton direction, we find that
$${\partial V \over \partial S} \propto \Omega_S-{\Omega \over 2S_R}
-\Omega(log[det A]^{1/N})_S, \eqno(33)$$
which means that $F_S=0$ in vacuum as in the pure gauge case.

The minimization condition in the $T$
direction now reads (defining $G^{\prime}(T)=G_2(T)-{3/ 2T_R d^{\prime}}
+{\phi_i \phi_i^{\dagger}/ 4 T_R^2 d^{\prime}}$)
$$\eqalignno{{\partial V \over \partial T}&={e^{-\phi_i \phi_i^{\dagger}/2T_R}
\over {16 S_R T_R^3}}
{|[det A]^{1/N}|^2  \over {|\eta(T)|^{8 \pi d^{\prime}}}}
\{d^{\prime} G_2^{\prime}[|2S_R \Omega_S-\Omega-{2 \Omega r \over N}|^2+
|\Omega|^2 \left({4 d^{\prime 2} T_R^2 \over 3} | G_2^{\prime}|^2-3\right)] \cr
&+{2 d^{\prime} T_R \over 3}|\Omega|^2[2| G_2^{\prime}|^2+T_R  G_2^{\prime*}
 G^{\prime}_{2T}+T_R  G_{2T}^{\prime*}  G_2^{\prime}] \}=0, &(34)}$$
compared to Eq. (19). Writing Eq. (34) we made the approximation
$3 T_R \sim 3T_R-\phi_i \phi_i^{\dagger}$ for simplicity. The numerical
analysis was performed for the exact potential without this simplification.
The first term in the curly brackets vanishes because
$F_S=0$ in the presence of hidden matter.

The maxima and the saddle points of $V$ which were given by the solution to
$$\hat G_2(T)= G_2(T)-{\pi \over T_R}=0, \eqno(35)$$
in the pure gauge case are now given by the solution to
$$G^{\prime}(T)=G_2(T)-{3 \over {2T_R d^{\prime}}}
+{\phi_i \phi_i^{\dagger} \over 4 T_R^2 d^{\prime}}=0. \eqno(36)$$
Note that $G^{\prime}(T)$ does not have a well--defined modular weight
(i.e. not a modular function like $\hat G_2(T)$) due to the VEVs of $\phi_i$
in the matter Kahler potential $K(\phi_i, \phi_i^{\dagger})$ and $det A$
which break target space duality spontaneously.
As a result, these points are no longer the fixed points of target space
duality but simply solutions of Eq. (36). Also, we see that the
location of the maxima and saddle points depend mainly on the parameter
$d^{\prime}$ and not on the string model parameters $N,M$ and $t$ separately.
In addition, there is a weak dependence on the scalar VEVs
$\l \phi_i \phi_i^{\dagger}\r$ (fixed to be $\sim 0.2$) which we have
numerically found to be not important.

We find that as $d^{\prime}$ decreases, i.e.
the hidden matter content ($M$) increases and/or the hidden masses decrease
($t$ increases), $T_R$ at the maxima and saddle
points increase. The maxima of the pure gauge case at $(1,n)$ are now at
$(T_{Rmax},n)$ where $T_{Rmax}$
is given in Table 1 for some values of $d^{\prime}$.
The saddle points which for the pure gauge case were at ($\sqrt 3/2,1/2+n$)
are now given by $(T_{Rsp},1/2+n)$ where $T_{Rsp}$ are given
for the same values of $d^{\prime}$ in Table 1. It is easy to see from
Table 1 that as there is more hidden matter (increasing $M$)
and/or the hidden masses become
smaller (larger $t$) the maxima and saddle points appear at larger values of
$T_R$. Note that the value of
$T_I$ depends on $d^{\prime}$ very weakly since the minimization
condition for $T_I$ (and not $T$ given by Eq. (34)) is almost $d^{\prime}$
independent because at the minimum $d^{\prime} T_R $ is (almost) constant.
We also see that the $T_I$ values are periodic with a period of 1
since the modular functions which appear in the effective scalar potential,
$\eta(T)$ and $G_2(T)$, are periodic functions of $T_I$ with the same period.

The minima of the pure gauge case which were at $(\sim 1.23,n)$ and
$(\sim 0.81,n)$ are now given by the solutions to Eq. (36) which are not
zeros of $G^{\prime}(T)$. Again the $T_R$
at the minimum mainly depends on $d^{\prime}$ rather than on $N,M$ and $t$
separately. A numerical
study of the scalar potential gives the values in Table 1 for the $T_{Rmin}$
for some values of $d^{\prime}$. We see that $T_{Rmin}$ increases with
decreasing $d^{\prime}$ but at small (i.e. $<1/7$) values of $d^{\prime}$
a new minimum with very small (i.e. $<0.2$) $T_R$ appears in addition to
the one with large $T_R$.
Once again, as for the maxima and saddle points, $T_I$ at the minima are
periodic and (almost) independent of
$d^{\prime}$. Now however, contrary to the pure gauge case,
the minima for a given $d^{\prime}$ are not connected to others by target
space duality transformations. This is because target space duality
is spontaneously broken by the VEVs of $\phi_i$ which results in a scalar
potential which does not have a well--defined modular weight.
Note also that the minimum for the case with no hidden matter, i.e. $M=t=0$,
in Table 1
does not reproduce the pure gauge result because of the Kahler potential
term for the scalars $\phi_i$ in the potential.

When $d^{\prime}\leq 0$, we find that there is no minimum in the modulus
direction, i.e. $T_R \to \infty$. The reason is the change of sign in the
exponential of $T_R$ which comes from $\eta(T)$.
In this case one also gets $F_T=0$ so there is no SUSY breaking in the modulus
direction. Unless $T_R$ is stabilized, one does not have a well-defined
vacuum and cannot obtain SUSY breaking.
This is a new stability problem in addition to the
one for the dilaton. Compared to the dilaton case,
this result is much more difficult to modify since the
modulus dependence of the nonpertirbative superpotential is strongly
constrained due to target space duality.
In order to avoid this situation, we require
$d^{\prime}>0$ or $6N-2M-t>0$ which severely constrains the hidden sectors
of possible realistic
models. Without hidden matter masses, i.e. $t=0$, $d^{\prime}>0$
always since this is required for the asymptotic freedom of the hidden sector.
For $M$ copies of hidden matter in the fundamental representation
$N+ \bar N$, using Eq. (26) for $det A$ and Eq. (2) for the matter mass terms,
we have $t=\sum_{i=1}^M (2n_i-6)$
where $n_i$ is the order at which the mass term appears in the superpotential.
Requiring that the hidden matter remains in the spectrum around the
condensation
scale $\Lambda_H$, we get $n \sim 7-8$ or larger thus giving $t \sim 10M$.
The condition for stability in the $T_R$ direction loosely becomes
$3N-5M >0$ which is a rather strong condition on the hidden sector of
realistic models. Of course, $M$ is the number of light hidden matter
multiplets and not their overall number. A given
string model cannot be ruled out on the basis of the massless string spectrum
using the above condition since some or all of the hidden matter can get
large masses and decouple due to matter VEVs. Once all hidden masses are
found though, the above condition must be satisfied in order to get realistic
SUSY breaking and a stable vacuum in the modulus direction.

We see that the minimum of the effective scalar potential is not at a fixed
point of target space duality neither for the pure gauge case (due to
$K(\phi_i, \phi_i^{\dagger})$) nor for the case with hidden matter
(due to $det A$). This is a result of the spontaneous breaking of target
space duality by the scalar VEVs. On the other hand, it is
well--known that free fermionic strings correspond to orbifold models
formulated at the fixed points of target space duality i.e. $T_R=1$
(in units of $M_v$). How should one interpret the above results obtained from
the low--energy effective field theory?
First note that the overall modulus $T$ in the supergravity basis that
appears in the low--energy scalar potential is related to the overall modulus
$t_s$ in the string basis by[\CVET,\NANO,\FONT]
$$t_s={{T_c-T} \over {\bar T_c+T}}, \eqno(37)$$
where $T_c=1$.
Thus at the fixed point of target space duality $T=1$ the VEV of the modulus
in the string basis vanishes, $t_s=0$. This is fine since  modulus field
$t_s$ is the coefficient of the exactly marginal operator
$$:y^i \omega^i:: \bar y^i \bar \omega^i: \quad i=1,\ldots,6 \eqno(38)$$
which deforms the original two dimensional free fermionic string action.
Here $y^i, \omega^i,
\bar y^i, \bar \omega^i$ are the internal world--sheet fermions which describe
the compactified six dimensional manifold in the fermionic
language[\MOD,\NANO].
For $T=1$, $t_s=0$ and one
has a free fermionic string as expected. Once $T \not=1$, we get a nonzero
$t_s$ and therefore the free fermionic string is deformed by the above
Abelian Thirring interaction.
Consequently, one should understand the result of this section as follows.
Hidden gaugino and matter condensation and the scalar VEVs produce an
effective scalar potential
which perturbs the initial value of the modulus to the values of $T_{Rmin}$
given in Table 1 for different values of the parameter $d^{\prime}$.
As a result, the free fermionic string is perturbed by the
above Abelian Thirring interaction with the coefficient given by the value of
the modulus in the string basis $t_s$. The low--energy model with the scalar
VEVs, hidden masses and hidden matter and gaugino condensates is not described
by the original free fermionic string but by one which is perturbed by the
corresponding exactly marginal operator.

\bigskip
\centerline{\bf 5. SUSY breaking by hidden matter condensation}

In this section we consider the second case mentioned above in which
$m_{\phi_i}<\Lambda_H$ and $\phi_i$ remain in the spectrum.
Then, $\phi_i$ should be
treated as dynamical fields similar to $S$ and $T$ since they can be excited
due to their small masses. Now $W=W(S,T,\phi_i)$
where from Eq. (24) all the $\phi_i$ dependence is in the term $det A$ which
arises due to the matter condensates $\Pi_{ij}$. In this case, in addition to
$F_S$ and $F_T$, one should also evaluate $F_{\phi_i}$ since it can
be nonzero in the vacuum resulting in SUSY breaking in the matter direction.
It may also be possible to break SUSY mainly by
hidden matter condensation in the matter direction rather than by hidden
gaugino condensation in the modulus direction, i.e. $F_{\phi_i}>F_T$ in vacuum.
We consider two cases depending on the Kahler potential of the matter
fields $\phi_i$ whose F--terms are nonzero.
First, we investigate the effective potential when the matter Kahler potential
depends on the modulus as given by Eq. (4). Then, we repeat the same analysis
for matter
with canonical Kahler potential. Both cases and a mixture of the two are
possible depending the details of the string model.

The hidden matter condensates, through the term $det A$, induce an F--term
in the matter direction, $\phi_i$
$$\eqalignno{F_{\phi_i}={e^{-{\phi_i \phi_i^\dagger}/4T_R} \over {(2S_R)^
{1/2} (2T_R)^{3/2}}} &[\Omega(S) h(T) [det A]^{1/N} \cr
& \times \left({s_i \over {N \phi_i}}+{\phi_i^\dagger \over
4T_R}\right)+(W_{3\phi_i}+K_{\phi_i}W_3)]. &(39)}$$
This is the result obtained in Ref. (\EA) where the effect of matter
condensation on $F_{\phi_i}$ due to hidden matter mass terms was examined.
The last two terms simply give the contribution coming from the cubic
superpotential which vanishes for the solution to the F and D constraints
in Eqs. (5a-c) before the hidden gauge group condenses.
Generically the F and D flat solutions give $\l \phi_i \r \sim M_v/10$, a
scale which is set by the coefficient of the anomalous D--term in Eq. (5a).
The nonperturbative scalar potential also contains the scalars $\phi_i$ but
since it is much smaller than the tree level potential we assume that it
does not modify the VEVs of $\phi_i$ appreciably. Therefore, we set the second
paranthesis above to zero.
We see that for realistic values of $s_i$ and $N$ (i.e. of the same order of
magnitude) the first term in the second
paranthesis in Eq. (39) (which corresponds to the $W_k$ piece in $F_k$)
dominates the second one when $\l \phi_i \r \sim M_v/10$ unless $T_R$ is
very small (i.e. $<0.01$). In order for these
two terms to cancel each other, one needs either $\l \phi_i \r \sim M_v$
and $T_R \sim 1$ or $\l \phi_i \r \sim M_v/10$ and $T_R \sim 0.01$ and a
considerable amount of fine tuning.
$F_{\phi_i}$ obviously arises solely from matter condensation since its
origin is the hidden matter mass term $tr A \Pi$ in Eq. (20).

In Ref. (\EA), it was shown that $ F_{\phi_i}$ may be nonzero in vacuum once
$W_n$ or hidden matter
mass terms are taken into account. The reason is that, the $n>3$ terms give
corrections to the cubic superpotential, $W_3$, which modify the cubic level
F constraints. For large orders $n$ these corrections turn the F constraints
into an inconsistent set of equations. As a result, the new
set of F constraints up to a given order $n>3$,
cannot be solved simultaneously for any set of scalar VEVs. In particular,
at the minimum, there is always a nonzero $F_{\phi_i}$ for some $\phi_i$
and SUSY is spontaneously broken
in the matter direction. The amount of SUSY breaking in the matter direction
given by $F_{\phi_i}$
depends on the parameters of the model such as $M,N,t$ and $s_i$.

We stress that $\phi_i$
are connected to the squarks and sleptons either through gravity
or the broken gauge $U(1)$s. (We neglect the nonrenormalizable interactions
which do not affect our results.)
Since the $U(1)$s are broken at the high scale
of $O(M_v/10)$, interactions of $\phi_i$ with the squarks and sleptons are
suppressed by $O(10/M_v)$ and thus are almost as weak as gravity. Therefore,
this mechanism has all the characteristics of hidden sector supersymmetry
breaking rather than visible sector supersymmetry breaking[\DINE].

Now, for $F_{\phi_i} \not=0$ (and $F_S=0$), when the Kahler potential of
$\phi_i$ is given by Eq. (4), the scalar potential becomes
$$\eqalignno{V&={e^{-\phi_i \phi_i^{\dagger}/2T_R} \over {16S_R T_R^3
|\eta(T)|^{8 \pi d^{\prime}}}}
|[det A]^{1/N}|^2 |\Omega|^2
\{-2T_R |{s_i \over {N \phi_i}}+{\phi_i^{\dagger} \over 4T_R}|^2
\cr
&+ \left({4 d^{\prime2}T_R^3 \over (3 T_R- \phi_i \phi_i^{\dagger})}
|\{G_2(T)-{3 \over {2T_R d^{\prime}}}+{\phi_i \phi_i^{\dagger} \over {4 T_R^2
d^{\prime}}}|^2-3 \right)\}. &(40)} $$
The minimum in the modulus direction is given by
$$\eqalignno{{\partial V \over \partial T}=&{e^{-\phi_i \phi_i^{\dagger}/2T_R}
\over {32 \pi S_R T_R^3}} {|[det A]^{1/N}|^2 |\Omega|^2 \over
{|\eta(T)|^{8 \pi d^{\prime}}}} \{d^{\prime} G_2^{\prime}
[|{s_i \over {N \phi_i}}+{\phi_i^{\dagger} \over 4T_R}|^2+
\left({4 d^{\prime 2} T_R^2 \over 3} | G_2^{\prime}|^2-3\right)] \cr
&-{|\phi_i|^2 \over {8T_R^2}}-|{s_i \over {N \phi_i}}+{\phi_i^{\dagger}
\over 4T_R}|^2
+{2 d^{\prime} T_R \over 3}|\Omega|^2[2| G_2^{\prime}|^2 \cr
&+T_R  G_2^{\prime*}
 G^{\prime}_{2T}+T_R  G_{2T}^{\prime*}  G_2^{\prime}] \}=0, &(41)}$$
where we used $3T_R \sim 3T_R - \phi_i \phi_i^{\dagger}$ in the above
expression for simplicity. We stress that the numerical minimization was
performed without this simplification.

Compared to the previous case, the nonzero $F_{\phi_i}$ modifies the
effective potential by the first term in the curly brackets in Eq. (40).
The behavior in the modulus direction is altered
significantly due to the modulus dependence of the matter Kahler potential.
A numerical analysis shows that the above potential does not have a minimum
in the $T_R$ direction (at least for realistic values of the model parameters).
Thus, there
is a stability problem in the modulus direction if $F_{\phi_i} \not =0$ and
the matter Kahler potential depends on the modulus. This stability problem
arisies because of the first term in the curly brackets in Eq. (40), i.e. the
modulus dependent matter Kahler potential. For realistic values of $F_{\phi_i}$
(which are given by $3<s_i/N \phi_i<30$) and other model parameters, this
term destroys the
minima we found for the $F_{\phi_i}=0$ case in the previous section.
This is easy to understand since the new term is given (for $T_R>0.5$ which is
true for all minima) by $-T_R$ times a large
number of $O(10)$ which is enough to eliminate the minima. In light of the
above result,
since we want a potential which is stable in the $T_R$ direction, we will
assume that $F_{\phi_i}=0$ for all matter fields $\phi_i$ with a modulus
dependent Kahler potential.

It seems that there cannot be a nonzero matter F--term which results in a
stable potential in the $T_R$ direction. This is not so as we will now show.
As we saw above, the source of the problem is the modulus dependent
matter Kahler potential. Therefore, the simplest solution
is to find matter with a Kahler potential which does not depend on the modulus.
In fact, there are matter fields (which we denote $\phi_i$
again for simplicity) with canonical Kahler potentials i.e.
$$K(\phi_i, \phi_i^{\dagger})=-\sum_i \phi_i \phi_i^{\dagger}. \eqno(42)$$
These are untwisted matter fields which arise from sectors with all their
moduli projected out due to the twists of the basis vectors which define
the free fermionic string model[\MOD]. For example, if there are four complex
world--sheet fermions,
there is one sector with no moduli and all matter fields arising from this
sector have canonical Kahler potentials. If there are six complex world--sheet
fermions all untwisted moduli are projected out. Then, all untwisted matter
fields have canonical Kahler potentials which is the optimal case.

In this optimal case, when there are no untwisted moduli, the matter
F--term becomes
$$\eqalignno{F_{\phi_i}={e^{-{\phi_i \phi_i^\dagger}} \over {(2S_R)^
{1/2} (2T_R)^{3/2}}} &[\Omega(S) h(T) [det A]^{1/N} \cr
& \times \left({s_i \over {N \phi_i}}+{\phi_i^\dagger \over
M_v^2}\right)+(W_{3\phi_i}+K_{\phi_i}W_3)]. &(43)}$$
The only difference between the above formula and Eq. (39) is in the
Kahler term in front
and the second term in the paranthesis which are independent of the
modulus. The effective scalar potential becomes
$$\eqalignno{V&={e^{-\phi_i \phi_i^{\dagger}} \over {16S_R T_R^3
|\eta(T)|^{8 \pi d^{\prime}}}} |[det A]^{1/N}|^2 |\Omega|^2
\{|{s_i \over {N \phi_i}}+{\phi_i^{\dagger} \over M_v^2}|^2 \cr
&+ \left({4 d^{\prime2}T_R^2 \over 3}
|G_2(T)-{3 \over {2T_R d^{\prime}}}|^2-3 \right)\} &(44)} $$

We see that, due to the modulus independent Kahler potential, the problematic
factor $(K_{\phi_i \phi_i^{\dagger}})^{-1}=-2T_R$ is absent in this case.
Now the minimum of $V$ depends on the value of $F_{\phi_i}$ in addition to
the parameter $d^{\prime}$. From Eq. (39) we see that this is fixed by
the combination $s_i/N \phi_i$. In Table 2, we give the minima of the scalar
potential for some values of $d^{\prime}$ and three realistic values of
$F_{\phi_i}$ given by $(s_i/N \phi_i)^2=10,10^2,10^3$. We find that
the location of the minima does not depend strongly on $s_i/N \phi_i$
between these values. Since $K_{\phi_i \phi_i^{\dagger}}$
and $F_{\phi_i}$ do not depend
on $T$ (except for the Kahler potential term), the effect of a nonzero
$F_{\phi_i}$ is to change the $-3$ term in the scalar potential to
$(s_i/N \phi_i)^2-3$. From Table 2, we see that similarly to the $F_{\phi_i}=0$
case, as $d^{\prime}$ decreases, i.e. as
there is more hidden matter and/or hidden matter becomes lighter, $T_R$ at
the minima increase. Also as expected, $T_I$ is periodic and depends on
$d^{\prime}$ very weakly. The main difference which is crucial
for our purposes is the
absence of the minima with small values of $T_R$ for small values of
$d^{\prime}$. Note that now there are no minima connected to the ones in the
table by target space duality, i.e. $T \to 1/T$ because the potential in
Eq. (44) does not have a well--defined modular weight.
{}From Table 2, we also see that the value of $T_R$ at the minima
for a given $d^{\prime}$ increases slowly with the value of $F_{\phi_i}$.
In addition, as in the $F_{\phi_i}=0$ case, for
$d^{\prime} \leq 0$ the scalar potential in Eq. (39) is unstable in the
$T_R$ direction. Therefore, we must require that the hidden sector satisfies
$d^{\prime}>0$ or loosely $3N-5M>0$ as before.

One can also have a mixed case, i.e. some of the matter fields ($\phi_i$)
have Kahler potentials which depend on moduli while the rest ($\psi_i$) have
canonical Kahler potentials. In this case, for reasons of stability in the
$T_R$ direction, we assume that $F_{\psi_i} \not =0$ whereas $F_{\phi_i}=0$.
The scalar potential is now given by
$$\eqalignno{V&={e^{-\phi_i \phi_i^{\dagger}/2T_R} \over {16S_R T_R^3
|\eta(T)|^{8 \pi d^{\prime}}}}
|[det A]^{1/N}|^2 |\Omega|^2
\{- |{s_i \over {N \psi_i}}+{\psi_i^{\dagger} \over M_v^2}|^2
\cr
&+ \left({4 d^{\prime2}T_R^3 \over (3 T_R- \phi_i \phi_i^{\dagger})}
|\{G_2(T)-{3 \over {2T_R d^{\prime}}}+{\phi_i \phi_i^{\dagger} \over {4 T_R^2
d^{\prime}}}|^2-3 \right)\}. &(45)} $$

The minima of the potential in the modulus direction are given in Table 3 for
three values of $s_i/N \psi_i$ as before. We find that the results are very
similar to the previous case given in Table 2. Once again as $d^{\prime}$
decreases $T_R$ at the minima increase with no small $T_R$ minima arising at
small $d^{\prime}$. $T_I$ is almost independent of $d^{\prime}$ and periodic as
before. As before, $T_R$ at the minimum (for a given $d^{\prime}$) slightly
increases with the value of the matter
F--term. The two cases give qualitatively the same results with small
quantitative differences which are not important. We conclude that the presence
of matter with modulus dependent Kahler potentials does not affect our
results as long as their F--terms vanish.

We find that, the scalar potential has stable minima when $F_{\phi_i} \not =0$
if $\phi_i$ are matter fields whose Kahler potential is independent of the
modulus. The presence of other matter with modulus dependent Kahler potentials
does not alter the qualitative results as long as they do not have nonzero
F--terms.  For our purposes the most important numerical result is
the absence of minima with $T_R<0.2$ for the $F_{\phi_i} \not =0$ case.
This will play an important role when we examine the direction of SUSY breaking
in field space.

In the presence of dynamical matter $\phi_i$, one can ask several questions.
First we see from Eq. (31) that there are two contrubutions to $F_T$; one from
gaugino condensation and the other from matter condensation. What are the
relative magnitudes of these? In particular, can the matter condensate
contribution dominate that of the gaugino condensate in $ F_T $?
(We remind that $ F_{\phi_i}$ arises solely from matter condensates.)
A simple analysis shows that the gaugino condensation contribution is
larger if $6N-2M-t>0$ and vice versa for all values of $T_R$ at the minimum.
Note that the two contributions enter $F_T$ with opposite signs.
We found above that unless this
condition holds there is no stable minimum in the $T_R$ direction and both
$F_T$ and $F_{\phi_i}$ vanish. Thus, in all cases with a stable minimum in
the $T_R$ direction, the hidden gaugino condensation contribution to $F_T$
is larger than that of the hidden matter condensation. For small values of
$6N-2M-t$, when there is a large number of hidden matter multiplets and/or
they are light, the two contributions are comparable. On the other hand,
for large values of $6N-2M-t$ the gaugino condensation contribution is
dominant.

The second and more important question is for what range of model parameters
$N,M,s_i,t$ etc. and vacuum (i.e. $T_R,T_I$)
does one of the F--terms dominate the other, i.e.
$F_T >> F_{\phi_i}$ or vice versa? The relative magnitude of these
two F--terms gives the direction of SUSY breaking in field space (assuming as
before that $F_S=0$).
If all matter $\phi_i$ have canonical Kahler potentials, from Eqs. (31) and
(39) we find the ratio
$${ F_T  \over  F_{\phi_i} } ={{N \phi_i}\over s_i}d^{\prime}
\left(G_2(T)-{3 \over {2T_R d^{\prime}}}\right), \eqno(46)$$
whereas for the mixed case we have
$${ F_T  \over  F_{\phi_i}} =\left({{4 N \phi_i T_R d^{\prime}} \over
{N |\phi_i|^2+4 T_R s_i}}\right)
\left(G_2(T)-{3 \over {2T_R d^{\prime}}}+{\phi_i \phi_i^{\dagger} \over {4
T_R^2
d^{\prime}}} \right). \eqno(47) $$

Using Eq. (46) for the case with only canonical Kahler potentials,
we find that for most values of $T_R$ at the minima in Table 2, $F_{\phi_i}>
F_T$. For example, for the two limiting values of $T_R$ in the table, i.e.
$T_R \sim 0.8$ and $T_R \sim 4.7$ we obtain $0.2$ and  $0.04$ for the
ratio $F_T/F_{\phi_i}$ (assuming $N= s_i$ for simplicity).
For smaller values of $d^{\prime}$ which gives larger
$T_R$ the ratio becomes even smaller. Of course, the ratio of F--terms also
depends on the value of $N/s_i$ which we took to be one. For example, at
$T_R \sim 0.8$ if $N >5 s_i$ one gets $F_T>F_{\phi_i}$. This becomes more
difficult when there is more and/or lighter hidden matter. Then, $d^{\prime}$
is small which gives large $T_R$ and this requires a
large ratio of $N/s_i$ which is very difficult (if not impossible) to realize.
For example, for $T_R>1$ one needs $N>10s_i$ or $N>10$ in order to get
$F_T \sim F_{\phi_i}$. On the other hand, the rank of the hidden gauge group
is $\leq 11$ which shows that this case is marginal whereas for larger $T_R$,
$F_T< F_{\phi_i}$ always.
It is only for small $T_R$ that $F_T$ can be naturally larger than
$F_{\phi_i}$. This is due to the very sharp increase (in absolute value) of
$G_2(T)$ with decreasing $T$. These
are exactly the minima which appear for small $d^{\prime}$
when $F_{\phi_i}=0$ as we saw in the
previous section. Now, however, when $F_{\phi_i}$ is nonvanishing, we find
that these minima disappear due to the modification of the $-3$ term
in the scalar potential by the $|s_i/N \phi_i|^2$ factor.
As a result, for a wide range of parameters
$N,M,s_i,t$ and $\phi_i \sim M_v/10$ we find that $F_{\phi_i}>F_T$.
Repeating the above analysis for the mixed case using Eq. (47) and the
results from Table 3 we get essentially identical results. This means that
the ratio of the F--terms, $F_T/F_{\phi_i}$, is not sensitive to the presence
of matter with modulus dependent Kahler potential as long as these do not
have nonzero F--terms.

The effect of the matter F--term is small only when $(s_i/N \phi_i)\sim 1$ and
then the $-3$ factor is not changed by much. This situation is similar to
the $F_{\phi_i}=0$ case and there are minima with small $T_R$ for small
$d^{\prime}$. From Eq. (46) we find that for $(s_i/N \phi_i)\sim 1$, $F_T$
is the dominant SUSY breaking effect if the vacuum is given by the minimum with
the small $T_R$ and not the large one. Assuming $\phi_i \sim M_v/10$, this
means that $N \sim 10 s_i$. For $F_T$ to be dominant, we need at least an
$SU(10)$ hidden gauge group and also that each scalar appear in the determinant
of the hidden matter mass matrix only once. As we remarked in the previous
paragraph, this is a marginal case at best which covers a small part of the
parameter space.

\bigskip
\centerline{\bf 6. Conclusions and discussion}

In this paper, we investigated the effects of hidden matter condensation
on SUSY breaking in SUGRA models derived from free fermionic strings. We
found that the location of the critical points of the effective scalar
potential depend mainly on one parameter, $d^{\prime}=(6N-2M-t)/
4 \pi$. Here $N,M$ and $t$ give the hidden gauge group, the number of hidden
matter multiplets and the power of $\eta(T)$ in $det A$ where $A$ is the hidden
matter mass matrix respectively. The other parameter which is given by the
VEVs of the scalar fields which give mass to hidden matter was taken to be
$\phi_i \phi_i^{\dagger} \sim 0.2$
since there are in general a large number of scalar VEVs with $O(1/10)$
(in units of $M_v$) fixed
by the anomalous D--term. We numerically checked that all our results
depend very weakly on the value of $\phi_i \phi_i^{\dagger}$ as long as it is
nonzero and in a realistic range.

When SUSY is not broken in the matter direction, we
found that as $d^{\prime}$ decreases, i.e. when there is more and/or lighter
hidden matter, $T_R$ at the maxima and saddle points of the effective
scalar potential increase such that $d^{\prime} T_R$ is about constant.
Contrary to the case of a hidden sector with a pure
gauge group, the minima and saddle points do not appear at the fixed points of
target space duality since this is spontaneously broken by the scalar VEVs
which give masses to hidden matter.
$T_R$ at the minima also increase with
decreasing $d^{\prime}$ but for values smaller than $\sim 1/7$ new minima with
small $T_R$ appear. $T_I$ at all the critical points are periodic since the
modular functions which enter the scalar potential are so. In addition,
$T_I$ at these points depend very weakly on $d^{\prime}$.
When $d^{\prime} \leq 0$, we found that there is no stable minimum in the $T_R$
direction, i.e. $T_R \to \infty$.
This stability problem is much more severe
than the one for the dilaton since the modulus dependence of the effective
nonperturbative superpotential is severely restricted by target space duality.
In order to avoid this problem, realistic models must satisfy $d^{\prime}>0$
or $6N-2M-t>0$ which is a strong restriction on their hidden sectors. One
cannot rule out string models on the basis their massless spectrum
since part or all of the hidden matter can get large masses from scalar VEVs.
Thus, the above condition should be used only for the part of the hidden sector
which does not decouple from the theory at the condensation scale.

When SUSY is also broken in the matter direction, i.e. $F_{\phi_i} \not =0$
in addition to $F_T \not=0$, the results depend on the Kahler potential of
the matter fields $\phi_i$. If $K(\phi_i, \phi_i^{\dagger})$ depends on moduli
there is no stable minimum for the scalar potential in the $T_R$ direction.
This is true for all matter fields which arise from sectors with moduli.
In order to get a stable potential, one must assume that all such matter fields
have vanishing F--terms.
This can happen if they do not enter the hidden mass matrix and therefore
have vanishing F--terms due to the cubic level constraints.
On the other hand, there are models in which some or all sectors are without
moduli. In that case, the Kahler potential of matter fields coming from these
sectors is canonical. We find that the scalar potential has a stable minimum
in the presence of nonzero matter F--terms if they correspond to fields with
canonical Kahler potentials. The presence of additional matter with modulus
dependent Kahler potentials does not destroy the stability as long as their
F--terms as zero. The behavior of the critical
points is very similar to the case with $F_{\phi_i}=0$ as is seen from the
Tables 1,2 and 3. We also find that the dependence of the minima on the value
of $F_{\phi_i}$ is weak.

When $F_{\phi_i} \not=0$, the most important numerical result for our purposes
is the absence of minima with small (i.e. $<0.2$) $T_R$. As a result of this,
we
find that for a wide range of model parameters $F_{\phi_i}>F_T$. Only for
$N>5s_i$ with very little hidden matter or for $N>10s_i$,
$F_T \geq F_{\phi_i}$. We conclude
that for most of the parameter space SUSY is mainly broken by hidden matter
condensation in the matter direction rather than by hidden gaugino
condensation in the modulus direction.

We saw that in the presence of many hidden matter multiplets with small masses,
the minima of the scalar potential are at large ($>1$) $T_R$.
Vacua with large $T_R$ are desirable
for obtaining large string threshold corrections[\THR] to the running
coupling constants
of the Standard Model gauge group. It is well--known that string unification
occurs around $10^{17}~GeV$ which is an order of magnitude larger than the
scale predicted by the minimal supersymmetric extension of the Standard Model.
This discrepancy can be eliminated without introducing extra states only by
having large string threshold corrections which require large $T_R$. As we
saw above, large values of $T_R$, which are very difficult to obtain without
hidden matter, occur naturally when there is hidden matter
which condenses. One can also turn this argument around and find the range
of $T_R$ required to get unification of coupling constants from string
threshold corrections. This will give the realistic range of $d^{\prime}$
which in turn gives possible values of the string model parameters $N,M,t$ etc.
However, a given value of $d^{\prime}$ does not fix the parameters since
there are different combinations of them which result in the same $d^{\prime}$.

At the TeV scale, the only way to find out the direction of SUSY breaking
in field space is to examine the sparticle masses (or soft--SUSY breaking
parameters in general). It is well--known that these exhibit distinct patterns
when SUSY is broken dominantly in the dilaton or the moduli directions[\SSB].
For the former the soft--SUSY breaking masses are all equal whereas for
the latter they depend on the modular weights of the observable fields and
in general are not equal to each other. One
can extend these ideas to SUSY breaking in the matter direction and find
the behavior of soft--SUSY breaking masses in this case. This would require
information about the dependence of observable matter Kahler potentials on
the fields $\phi_i$. For observable matter coming from the twisted sectors
the form of the Kahler potential has been conjectured to have a simple
dependence on $\phi_i$[\KAH]. It is therefore plausible that SUSY
breaking by hidden matter condensation in the matter direction
leads to a pattern of sparticle masses which differs from
the other two. In that case, one would be able to look for signs of this SUSY
breaking mechanism around the TeV scale.

\bigskip
\centerline{\bf Acknowledgements}
I would like to thank Lance Dixon and Scott Thomas for useful discussions.
This work was supported by the Department of Particle Physics and a Feinberg
Fellowship.

\vfill
\eject

\refout
\vfill
\eject

\end
\bye

\input tables.tex
\nopagenumbers
\magnification=1200
\baselineskip=18pt
\hbox
{\hfill
{\begintable
\ &$d^{\prime}=(6N-2M-t)/4 \pi N$   \ \|\ $T_{max}$      \ \|\ $T_{sp}$   \
\|\ $T_{min}$   \crthick
          &$3/2 \pi$             \ \|\  (0.77,0.5+n)   \ \|\  (0.97,n)   \
\|\   (1.11,n)      \nr
          &$3/4 \pi$             \ \|\  (1.87,0.5+n)   \ \|\  (1.88,n)   \
\|\   (2.47,n)      \nr
          &$1/2 \pi$             \ \|\  (2.83,0.5+n)   \ \|\  (2.83,n)   \
\|\   (3.75,n)      \nr
          &$3/8 \pi$             \ \|\  (3.79,0.5+n)   \ \|\  (3.78,n)   \
\|\   (5.02,n)      \nr
          &$3/8 \pi$             \ \|\  (3.79,0.5+n)   \ \|\  (3.78,n)   \
\|\   (0.17,0.31)   \nr
          &$3/ 10 \pi$           \ \|\  (4.74,0.5+n)   \ \|\  (4.74,n)   \
\|\   (5.00,n)
 \endtable}
\hfill}
\bigskip
\parindent=0pt
\hangindent=39pt\hangafter=1

 Table 1.
Location of the maxima $T_{max}$, saddle points $T_{sp}$, and minima $T_{min}$
of the effective scalar potential for different values of the parameter
$d^{\prime}$ assuming that there is no matter F--term. $n$ is an integer.
\vfill
\eject

\end
\bye

\input tables.tex
\nopagenumbers
\magnification=1200
\baselineskip=18pt
\hbox
{\hfill
{\begintable
\ $d^{\prime}=(6N-2M-t)/4 \pi N   \ \|\ $T_{min}$ (1) \ \|\ $T_{min} (2)$
\ \|\ $T_{min}$ (3)\crthick
    3/2 \pi   \|\ (0.80,0.5+n)  \|\  (0.86,0.5+n) \|\ (0.87,0.5+n) \nr
    3/4 \pi   \|\ (1.74,n)      \|\  (1.89,0.5+n) \|\ (1.91,0.5+n) \nr
    1/2 \pi   \|\ (2.61,n)      \|\  (2.84,0.5+n) \|\ (2.86,0.5+n) \nr
    3/8 \pi   \|\ (3.48,n)      \|\  (3.78,0.5+n) \|\ (3.82,0.5+n) \nr
    3/ 10 \pi \|\ (4.35,n)      \|\  (4.73,0.5+n) \|\ (4.77,0.5+n)
 \endtable}
\hfill}
\bigskip
\parindent=0pt
\hangindent=39pt\hangafter=1

 Table 2.
Location of the minima $T_{min}$ of the scalar potential for different
values of $d^{\prime}$ and different values of matter F--term when the matter
Kahler potential does not depend on the modulus. The three
cases (1), (2), (3) in the table correspond to $|s_i/N \phi_i|^2=10,10^2,10^3$
respectively. $n$ is an integer.

\vfill
\eject

\end
\bye

\input tables.tex
\nopagenumbers
\magnification=1200
\baselineskip=18pt
\hbox
{\hfill
{\begintable
\ $d^{\prime}=(6N-2M-t)/4 \pi N   \ \|\ $T_{min}$ (1) \ \|\ $T_{min}$ (2)
\ \|\ $T_{min}$ (3)\crthick
    3/2 \pi   \|\ (0.76,0.5+n)  \|\  (0.77,0.5+n) \|\ (0.77,0.5+n) \nr
    3/4 \pi   \|\ (1.68,n)      \|\  (1.86,0.5+n) \|\ (1.87,0.5+n) \nr
    1/2 \pi   \|\ (2.54,n)      \|\  (2.81,0.5+n) \|\ (2.83,0.5+n) \nr
    3/8 \pi   \|\ (3.40,n)      \|\  (3.75,0.5+n) \|\ (3.78,0.5+n) \nr
    3/ 10 \pi \|\ (4.27,n)      \|\  (4.69,0.5+n) \|\ (4.74,0.5+n)
 \endtable}
\hfill}
\bigskip
\parindent=0pt
\hangindent=39pt\hangafter=1

 Table 3.
Location of the minima $T_{min}$ of the scalar potential for different
values of $d^{\prime}$ and different values of matter F--term for the mixed
case. The three
cases (1), (2), (3) in the table correspond to $|s_i/N \phi_i|^2=10,10^2,10^3$
respectively. $n$ is an integer.

\vfill
\eject

\end
\bye